\documentclass[11pt]{article}
\usepackage{sao1,graphics}

\newcommand{\te}{$T_{\rm eff}$}
\newcommand{\tev}[1]{$T_{\rm eff}=#1$~K}

\newcommand{\bsv}[1]{$\langle B \rangle=#1$~kG}
\newcommand{\st}{{\tt SYNTH3}}
\newcommand{\sm}{{\tt SYNTHMAG}}

\newcommand{\I}{\mbox{{\boldmath $I$}}}
\newcommand{\J}{\mbox{{\boldmath $J$}}}
\newcommand{\K}{\mathcal{K}}
\newcommand{\ion}[2]{#1~{\sc #2}}

\newcommand{\beq}{\begin{equation}}
\newcommand{\eeq}{\end{equation}}

\newcommand{\fifps}[2]{\centering\resizebox{#1}{!}{\includegraphics{#2}}}
\newcommand{\filps}[3]{\centering\resizebox{#1}{#2}{\includegraphics{#3}}}

\begin{document}

\title{Spectrum synthesis for magnetic, chemically stratified stellar 
atmospheres}
\author{O. Kochukhov}
\institute{Department of Astronomy and Space Physics, Uppsala University,
Box 515, SE-751 20 Uppsala, Sweden\\{\tt oleg@astro.uu.se}}

\maketitle 

\begin{abstract}
Modern investigations of magnetic chemically peculiar stars reveal a variety of
complex processes in their atmospheres. Realistic spectrum synthesis modelling
of these objects has to take into account anomalous chemical composition,
strong magnetic field and chemical stratification. These effects complicate
calculation of theoretical spectra, especially when one has to deal with large
number of lines and wide spectral regions. To overcome the formidable problem 
of comparing model and observed spectra for magnetic chemically peculiar stars,
a new suite of spectrum synthesis programs was developed.
Here we describe in detail the synthesis codes, \st\ and \sm\ and present examples of
their application to various aspects of the peculiar-star surface phenomena.
The new codes proved to be reliable tools for the line identification,
magnetic field determination, chemical abundance and stratification analysis.
\end{abstract}

\section{Introduction}

Recent major improvements in the quality of spectroscopic observations
stimulated a surge of interest in detailed model atmosphere and chemical
abundance studies of magnetic stars. Modern spectrographs, such as UVES at the
ESO 8-m VLT at Paranal, HARPS at the ESO 3.6-m telescope at La Silla, or the
NES instrument at the SAO 6-m telescope, can now provide $S/N>300$, wide
spectral coverage echelle spectra of moderately faint stars at the  resolving
power of $\lambda/\Delta\lambda=4\times10^4$--$10^5$. Publicly accessible data archives assist in
distributing these spectra to an astronomical community far wider than
the co-investigators on the original observing proposals or scientists
from the countries that own the telescopes and instruments.

Availability of the new observational material has to be matched by the
corresponding development of the new analysis techniques, capable of handling
large spectral regions and large number of lines without compromising accuracy.
The complex chemistry of chemically peculiar (CP) stars, coupled with the prominence of the magnetic
field and chemical stratification effects, implies that any realistic analysis of
the atmospheres of these stars must be partly or entirely based on the method
of spectrum synthesis. Moreover, to deduce accurate chemical abundances,
theoretical computations must properly account for the transfer of polarized
radiation through the chemically inhomogeneous stellar atmosphere. Numerical
schemes  implemented in the codes have to be stable against rapid variation of
the equilibrium model structure and opacity, associated with the chemical clouds,
located at different atmospheric heights.

Several computer codes for modelling the time series of Stokes parameter
spectra were developed in the past (Wade et al. 2001), but they are typically
constrained to the analysis of short wavelength regions and are not thoroughly
tested with the chemically stratified atmospheres. The lack of general-purpose
spectrum synthesis codes, suitable for modelling magnetic, chemically
stratified stellar atmospheres, stimulated development of new tools and major
improvement of the existing, widely used software. In this paper we describe the
current status and recent updates of our two  spectrum synthesis codes: \sm\
(magnetic spectrum synthesis) and \st\ (computation of non-magnetic stellar
spectra). The former code represents a major revision and improvement of the
early magnetic spectrum synthesis code by Piskunov (1999), whereas the latter
supersedes the well-known {\tt SYNTH} code (Piskunov 1992).

\section{Physical foundations}

\sm\ and \st\ are designed to calculate spectra emerging from the static, plane-parallel,
one-dimensional model atmosphere. The Local Thermodynamical Equilibrium
assumption is used throughout. These approximations are adequate for a wide
range of stars in the Main Sequence band. In practice, the codes were 
successfully used to model stars in the range of spectral classes from early-B to late-M.

\subsection{Opacities}

The polarized radiation is fully characterized by the Stokes vector
$\I=\{I, Q, U, V\}^\dagger$, where the Stokes parameters are defined according to
Shurcliff (1962). The \sm\ code solves the polarized radiative transfer equation in the form
\beq
\mu\frac{d\I}{dm} = -\K\I + \J,
\label{eq1}
\eeq
where $m$ denotes the column mass scale, which is used for tabulation of model 
atmospheres, and $\mu\equiv\cos{\theta}$ is the cosine of the angle between line of sight and
the local surface normal.
For the sake of brevity here we omit complete expressions of the elements of the
absorption matrix $\K$ and of the emission vector $\J$, since this was discussed
in detail by Piskunov \& Kochukhov (2002). 

In calculating the metal line opacities, \sm\ takes into account Zeeman or
Paschen-Back splitting patterns specified in the input line list. The absorption and
anomalous dispersion profiles for each Zeeman component are described by the Voigt
and Faraday-Voigt function, respectively. Radiative, van der Waals and Stark
broadening is calculated according to the damping parameters provided in the line
list or using classical expressions (Gray 1992). The chemical and ionization equilibrium 
is calculated for a prescribed
vertical distribution of abundances. Absorption profiles of all lines
contributing to a given wavelength are summed and multiplied by the trigonometric
functions of the angles $\gamma$ and $\chi$, which define orientation of the local
field vector with respect to the line of sight. The resultant coefficients enter
the absorption matrix $\K$.

The hydrogen line opacity is determined with the routines developed by Barklem et
al. (2000). The Stark broadening tables of St\'ehle (1994) are used to account for
the linear  Stark effect. Self-broadening of hydrogen lines (Barklem et al. 2000)
is also included. An optional module permits improved calculations of the
overlapping  lines of the higher hydrogen series members with the occupation
probability formalism of Daeppen et al. (1987) and Hubeny et al. (1994).

The Stark broadening parameters of the neutral He lines are computed following 
Dimitrijevic \& Sahal-Brechot (1984) and Freudenstein \& Cooper (1978).

The continuous opacity calculation in our spectrum synthesis codes is based upon the
routines adapted from the {\tt ATLAS9} code (Kurucz 1993). We include absorption due to
the bound-free and free-free transitions of \ion{H}{i}, H$_{2}^{+}$, H$^-$,
\ion{He}{i}, \ion{He}{ii}, different metal ions, He$^-$ free-free, Rayleigh
scattering  for \ion{H}{i} and \ion{He}{i}, and electron scattering.

Opacity calculations in \st\ are similar to those of \sm, except that a non-magnetic
version of Eq.~(\ref{eq1}) is used and unpolarized radiation is computed using the
usual scalar opacity coefficients. It was verified that the line profiles of \st\ 
are identical to calculations with \sm\ if magnetic field strength is set to zero.

\subsection{Ionization and molecular equilibrium}

Ionization and molecular equilibrium in \st\ and \sm\ is obtained with the advanced
equation of state solver written by N. Piskunov. This code represents a  modified and expanded
version of the routine described in Valenti et al. (1998). The solver includes the treatment of
$\approx200$ diatomic and polyatomic molecules and has been thoroughly  tested against similar
routines of the {\tt MARCS} and {\tt PHOENIX} model atmosphere codes for temperatures from
$\sim10^5$~K down to a few hundred K. 

Atomic partition functions are calculated using the tables given in the PFSAHA
subroutine of the {\tt ATLAS9} code (Kurucz 1993). For a number of heavy elements,
especially rare-earths, updated partition functions of Cowley \& Barisciano (1994)
and Sneden (2002) are incorporated. For calculations of the ionic populations we
typically take into account 6 ionization stages for light elements, 4 stages for the iron-peak
group and 3 stages for heavy elements. 

For diatomic molecules the partition function information is taken from Sauval \&
Tatum (1984) and Irwin (1987). For polyatomic molecules, the data from Irwin (1988)
are used.

\subsection{Magnetic field}

\begin{figure*}[t]
\fifps{8.5cm}{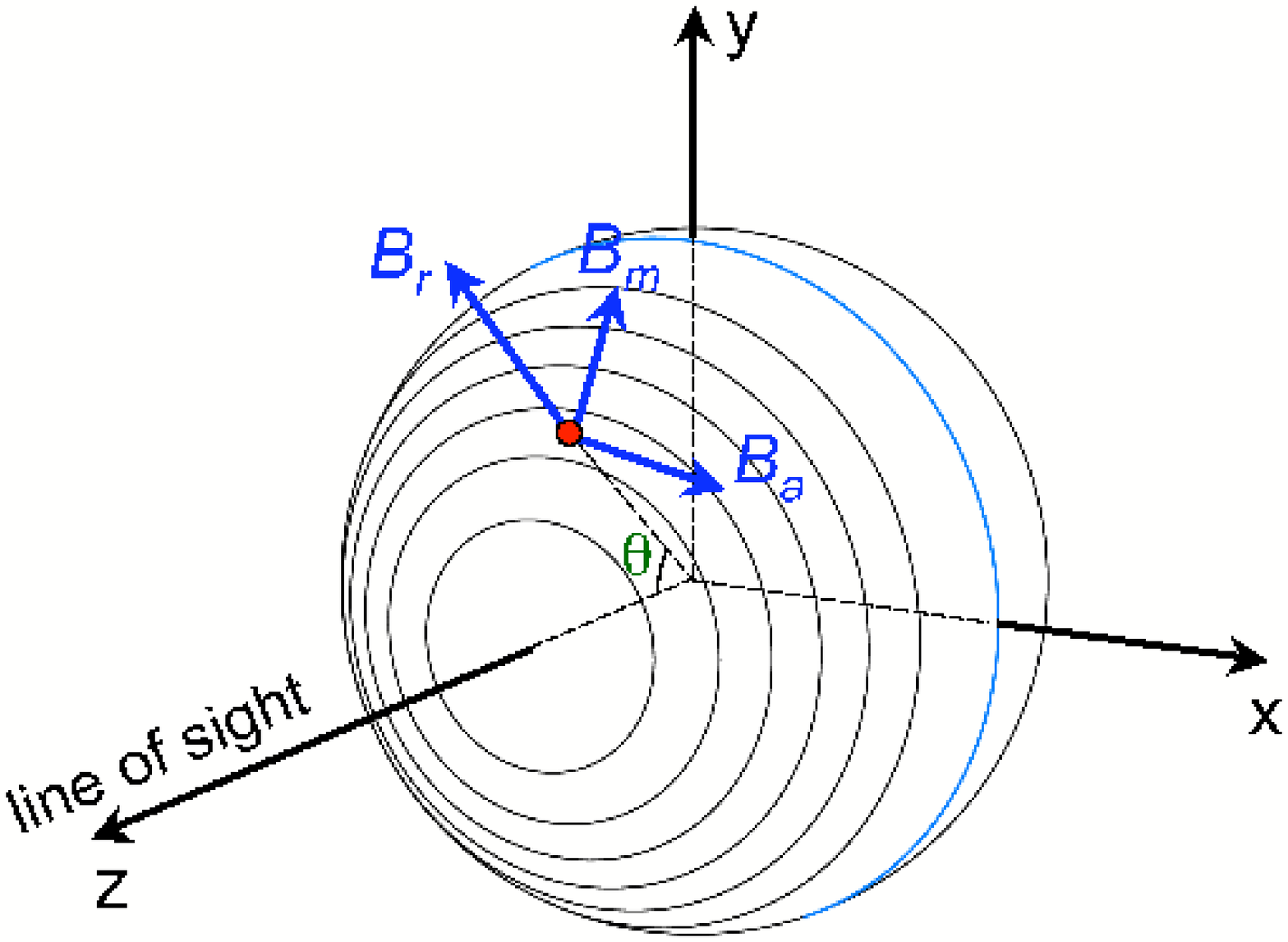}
\fifps{7cm}{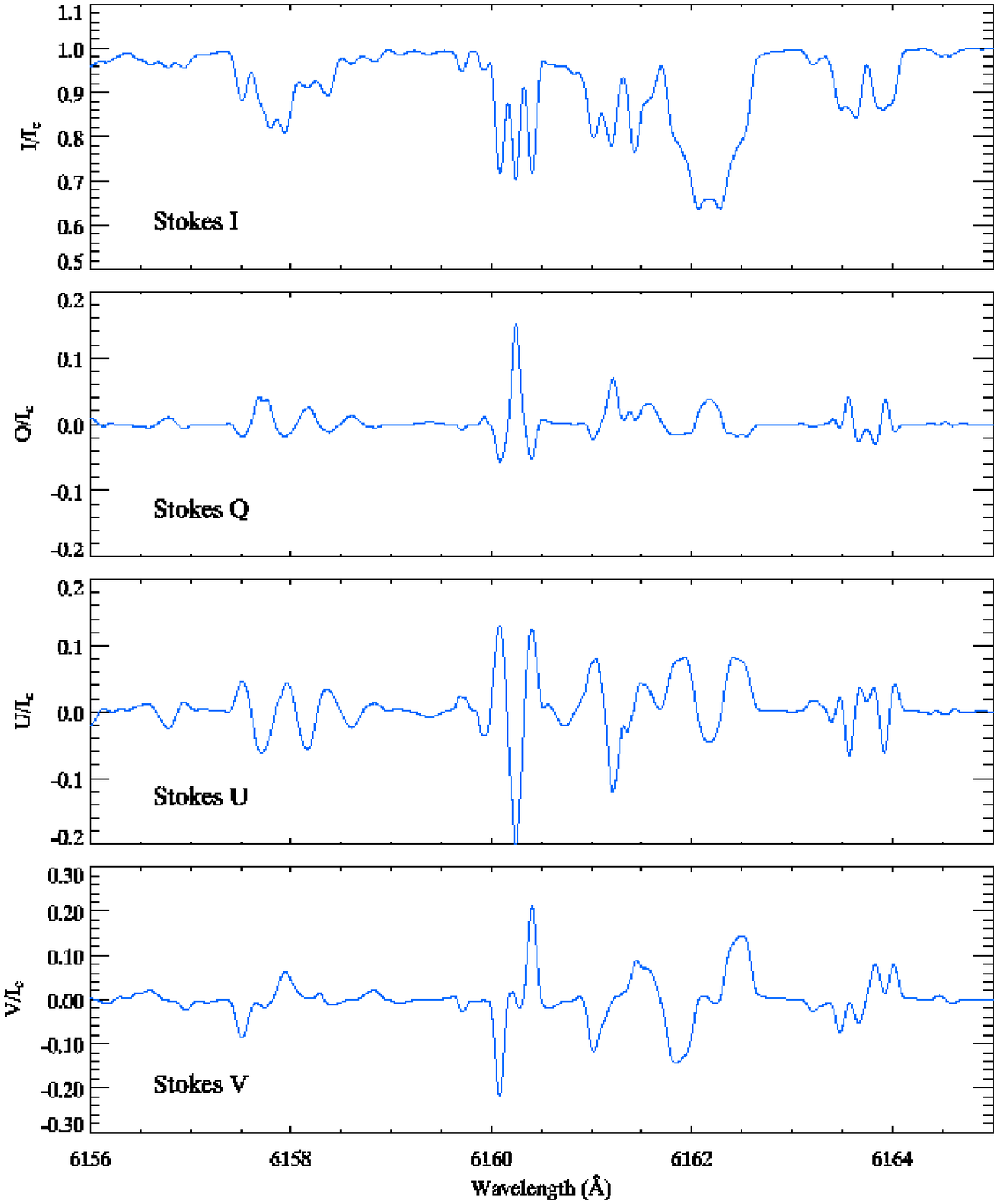}
\caption{{\it Left panel}: magnetic field geometry adopted in \sm. The field structure is
symmetric with respect to the line of sight ($z$-axis). The three magnetic vector
components, $B_r$, $B_a$, $B_m$, describe the radial, meridional and azimuthal
field in the stellar reference frame. Transformation of the local field vector
to the observer's coordinate system $xyz$ depends on the angle $\theta$ of each
of the seven annular zones employed for calculation of the disk-integrated profiles.
{\it Right panel}: typical four Stokes parameter disk-center \sm\
calculations for $B=10$~kG and \tev{7700}.}
\label{fig1}
\end{figure*}

Detailed structure of the stellar magnetic field geometry is best studied with the
time-resolved spectropolarimetric observations using Doppler imaging (Piskunov \&
Kochukhov 2002) or fitting parameters of the multipolar field models (Khalack \& Wade
2006). On the other hand, the main purpose of \sm\ is to properly account for the Zeeman
broadening and splitting in a wide wavelength region analysis of the unpolarized 
spectra of magnetic stars. Information content of a few unpolarized stellar
observations, available for the majority of CP stars, is usually insufficient to
constrain even the most simple model of the field geometry, especially when no
rotational modulation can be detected. This is why the common approach of expanding
the field structure in spherical harmonics is poorly constrained and one has to
resort to a simplified field model, characterized by a small number of free
parameters. Following this ideology, the \sm\ code calculates the local  Stokes
$IQUV$ spectra for a set of angles between the line of sight and local surface
normal assuming that i) the field is axisymmetric with respect to the line of sight
and ii) the field structure is homogeneous and is defined by the three vector
components: radial, meridional and azimuthal field (see Fig.~\ref{fig1}). This is
essentially equivalent to approximating the field strength distribution with a
single value -- an approach which works surprisingly well for many slowly rotating
CP stars.

\section{Numerical details}

Both the \sm\ and \st\ programs are written in Fortran 77, which ensures portability
to all major computer platforms and straightforward interfacing with alternative
software modules. The codes use standard BLAS and LAPACK routines, but do not depend
on commercial or proprietary software libraries. Up to now, the  codes were compiled
and successfully used on the SunOS and HP-UX versions of the Unix operating system, on
Mac OSX and Linux.

In the standard mode, \st\ computes intensity by solving numerically the scalar
radiative transfer equation for a set of (typically) seven $\mu$ angles. \sm\
proceeds similarly, but solves the vector polarized RT equation and derives Stokes
$IQUV$ profiles for a given local magnetic field.

Each code does the following set of operations:
\begin{itemize}
\item Reading the input model atmosphere (krz format) and atomic/molecular
line list in the VALD format (Kupka et al. 1999). For \sm\ the input line list also contains 
information on the strength and splitting of the Zeeman components for each
line. Depth dependence of chemical abundance, magnetic field strength or
microturbulent velocity can be specified in the input model atmosphere
file.
\item Concentrations of ions and molecules are computed with the equation of state
module described above. For each layer in the model atmosphere we calculate
continuous and line center opacities and the Voigt function parameters. The depth grid
is refined if necessary.
\item Intensity at several limb
angles is computed with the help of formal polarized (or scalar for \st) radiative 
transfer algorithm, using precomputed line-center opacities. The wavelength grid is
refined to obtain an accurate description of the line profile shapes. 
\item Intensity profiles are broadened for a given projected rotational velocity and 
macroturbulence. The resultant profiles are combined to form the disk-integrated spectra.
\end{itemize}

\subsection{Integration of the radiative transfer equation}

The polarized radiative transfer equation (1) is solved on discrete vertical grid
using the Diagonal Element Lambda Operator (DELO) formal solver, introduced by  Rees
et al. (1989). In our spectrum synthesis codes we employ a modified DELO algorithm, 
which utilizes a parabolic approximation of the source function (Socas-Navarro et
al. 2000). Piskunov \& Kochukhov (2002) have compared several widely used
polarized radiative transfer algorithms and found that the quadratic DELO method is
superior in terms of speed and accuracy to the original DELO solver and to all other
polarized radiative transfer schemes. For typical {\tt ATLAS9} model atmospheres of the Main 
Sequence B-F stars the accuracy of Stokes parameter calculation with the DELO algorithm 
is $\sim10^{-4}$.

\begin{figure*}[t]
\filps{12cm}{4.5cm}{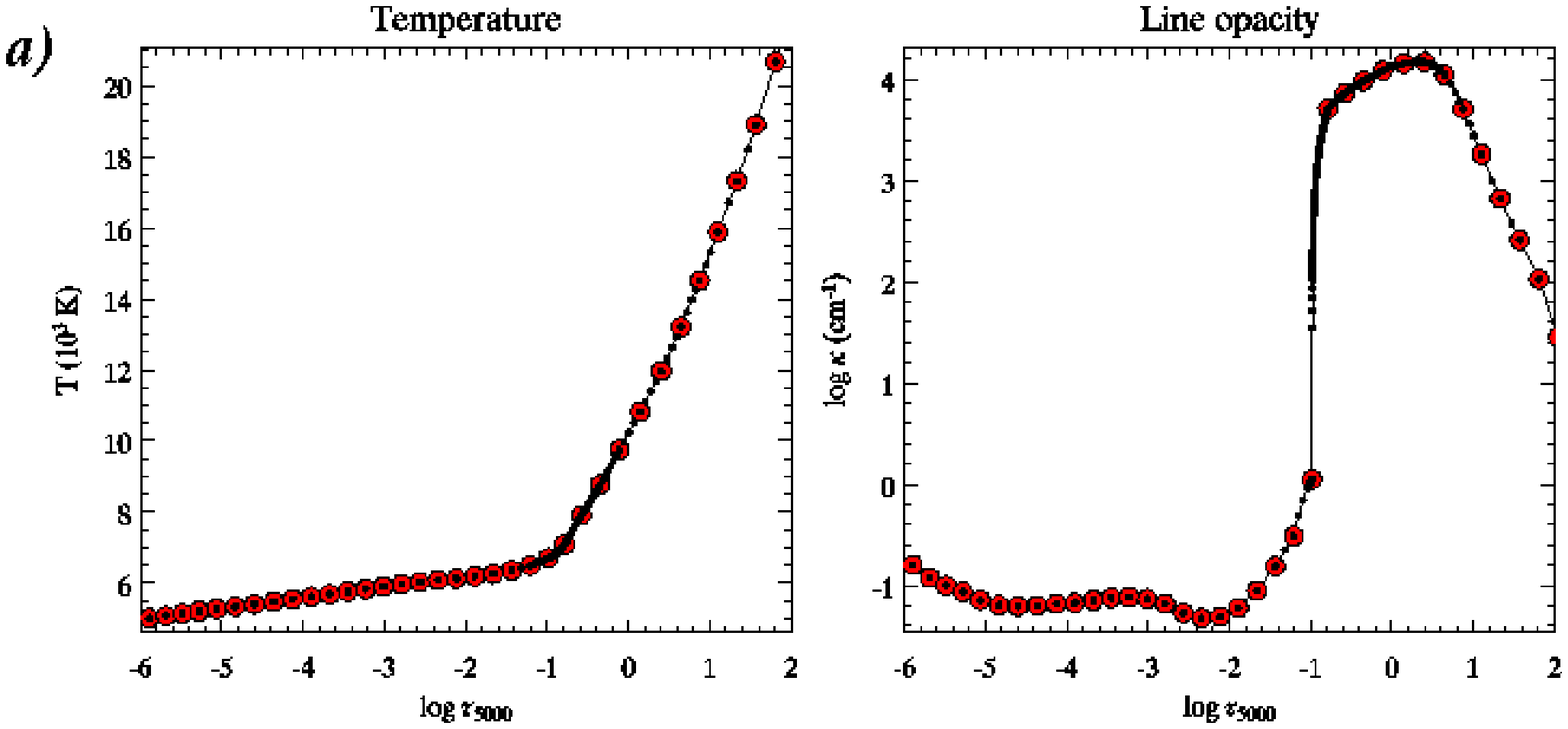}\\
\filps{12cm}{5cm}{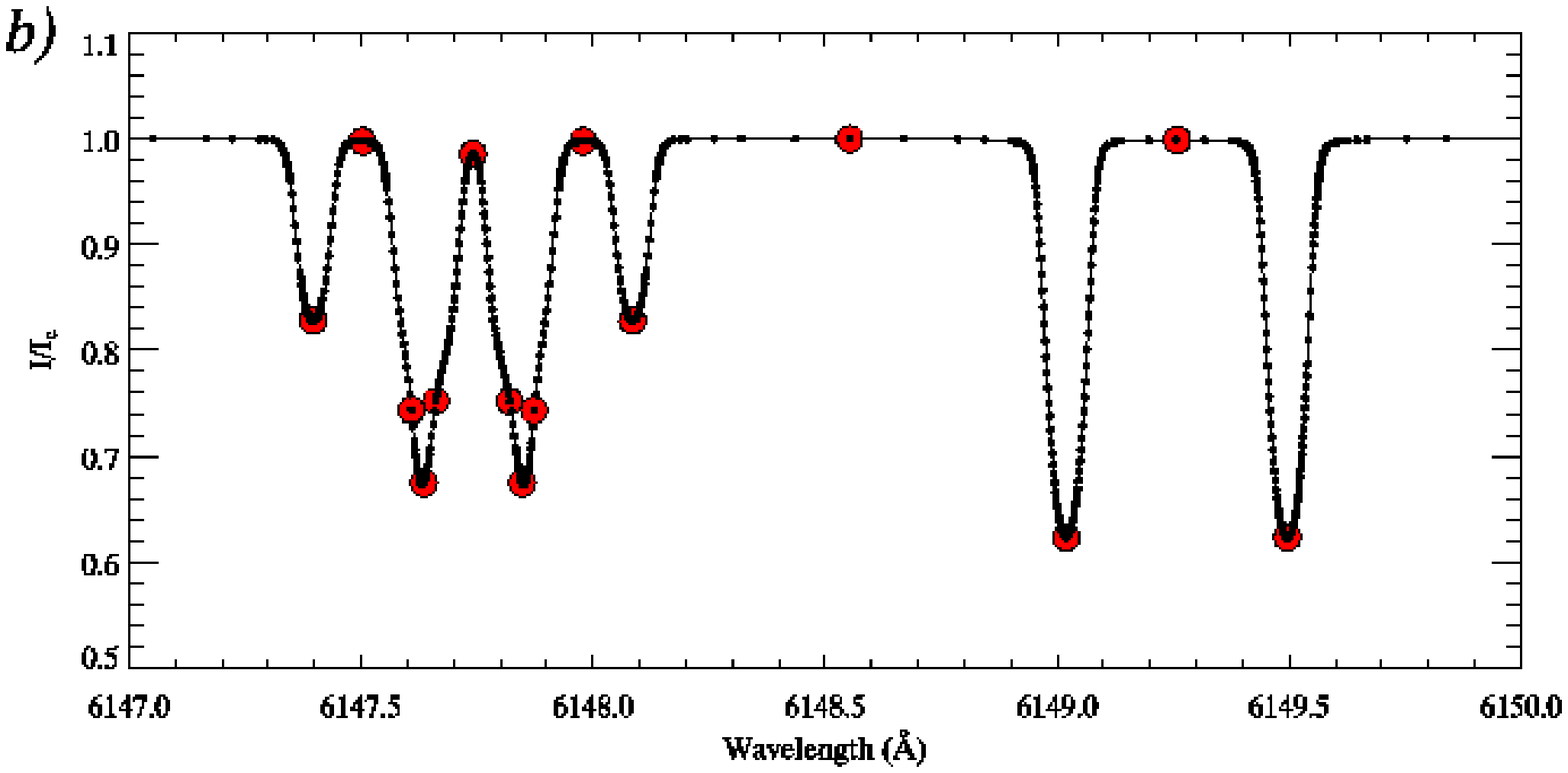}
\caption{Illustration of the adaptive grid refinement in \sm. {\bf a)} Extra
layers are added in the model atmosphere if vertical variation of the total opacity
is too fast. Large symbols show the original discretization of the model;
small symbols illustrate the final depth grid. {\bf b)} Refinement of
the frequency grid. Initially spectra are calculated for the centers of Zeeman
components and for the wavelengths in between (large symbols). More points (small symbols)
are added until the line profile shapes are described with sufficient
accuracy.}
\label{fig2}
\end{figure*}

Rapid variation of abundances with height in the atmospheres of chemically peculiar
stars poses further challenges to a radiative transfer solver. The latter has to provide 
accurate spectra for minimum number of layers in the model atmospheres and, at the
same time, be robust against steep variation of the line opacity resulting from
chemical  stratification. In the recent paper Kochukhov et al. (2006) investigated
performance of different scalar radiative transfer algorithms for chemically stratified
stellar atmospheres. These numerical experiments showed that direct intensity 
integration implemented in Kurucz (1993) codes is remarkably stable and performs well
for atmospheres with chemical stratification. However, this algorithm cannot be extended to
polarized radiation and for this reason it is not used in our codes. The quadratic
DELO scheme achieves acceptable accuracy of $\sim10^{-3}$ for the atmospheres with large abundance gradients
if extra vertical grid points are added between the layers where
opacity changes rapidly.

\subsection{Spectrum synthesis with adaptive grids}

To achieve required precision with the least amount of computations, adaptive grids
are extensively used in \sm\ and \st. At the stage of precalculation of the line-center
opacities, both codes consider the vertical gradient of the monochromatic optical depth.
If variation of the optical depth exceeds a given threshold, an extra layer is added
in the model structure. Tabulated model atmosphere quantities are obtained for additional
layers with the help of spline interpolation. The procedure of the depth grid refinement is
essential to obtain accurate spectra for the chemically stratified atmospheres. An example
of the depth grid refinement for cool Ap-star atmosphere with a strong vertical gradient of
the iron abundance is presented in Fig.~2a.

Most spectrum synthesis codes calculate line profiles on a wavelength grid defined by a fixed
wavelength step. This leads to computation of excessive number of continuum points with no
useful information. On the other hand, a coarse equidistant grid may not allow one to resolve details of
the line profile shapes. \sm\ and \st\ avoid this difficult compromise between the line profile
accuracy and  amount of computations by using adaptive wavelength grid. At the first step,
intensity is calculated at the centers of spectral lines or Zeeman components, if the magnetic
field is present. Then spectral points between each pair of the original wavelengths are
synthesized. Computed intensity is compared with the results of linear interpolation and
further frequency points are added if the discrepancy between interpolated and computed
spectra exceeds a given threshold. In this way, more points are calculated for the regions
with rapid variation of opacity, whereas the wavelength step remains relatively
large for continuum regions (see Fig~2b). Using this procedure, we obtain theoretical spectra
at infinite resolution, performing a factor of 10 to 50 less computations to achieve the same 
accuracy as in the synthesis with a sufficiently fine equidistant wavelength grid.

\subsection{Disk integration}

The final step of the spectrum synthesis calculation with \sm\ and \st\ is
integration of the local Stokes parameter/intensity spectra to obtain stellar flux
profiles. We use an external procedure, compatible with both the \sm\ and \st\
output files, to perform the disk integration. 

In the default mode, the local intensity and Stokes parameter spectra are calculated
for seven annular surface zones, chosen in such a way that their projected surface
areas are equal (see left panel in Fig.~1). Numerical experiments have demonstrated that this discretization of the
stellar surface is sufficient for computation of high-precision unpolarized
spectra. On the other hand, due to substantial variation of the shape and intensity
of the Stokes $QUV$ local profiles over the stellar surface in any realistic
magnetic field configuration, co-addition of seven linear and circular polarization
local spectra provides only a rough estimate of the signal expected in the $QUV$
disk-integrated spectra.

Intensity spectra are convolved with the appropriate rotational and  radial-tangential
macroturbulence broadening profiles (Gray 1992) and then added with equal weights.
This assumes spherically symmetric star, covered by a homogeneous magnetic field.
We note, however, that \sm\ and \st\ are, in fact, independent of the latter assumptions of the
disk integration and can be used to supply intensity  spectra for a more
sophisticated disk integration procedure. For instance, different sets of intensity
spectra can be combined to simulate observations of rotationally  distorted stars,
stars covered with chemical, temperature or magnetic spots, etc. Below we give an
example of such an advanced usage of \sm\ for calculation of the Zeeman resolved
spectral lines corresponding to dipolar field geometry.

\subsection{Reconstruction of chemical stratification}

\begin{figure*}[!th]
\fifps{13cm}{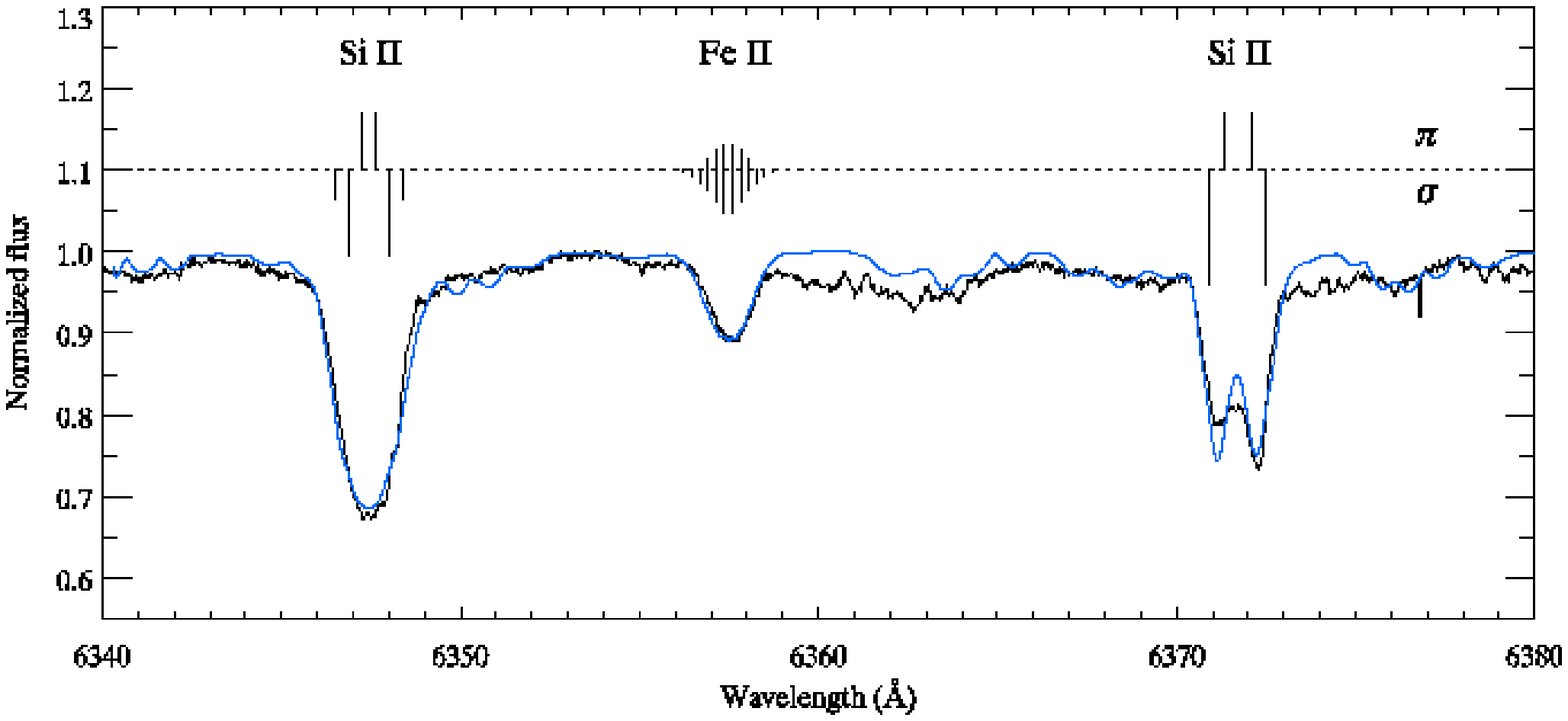}\\
\fifps{7.5cm}{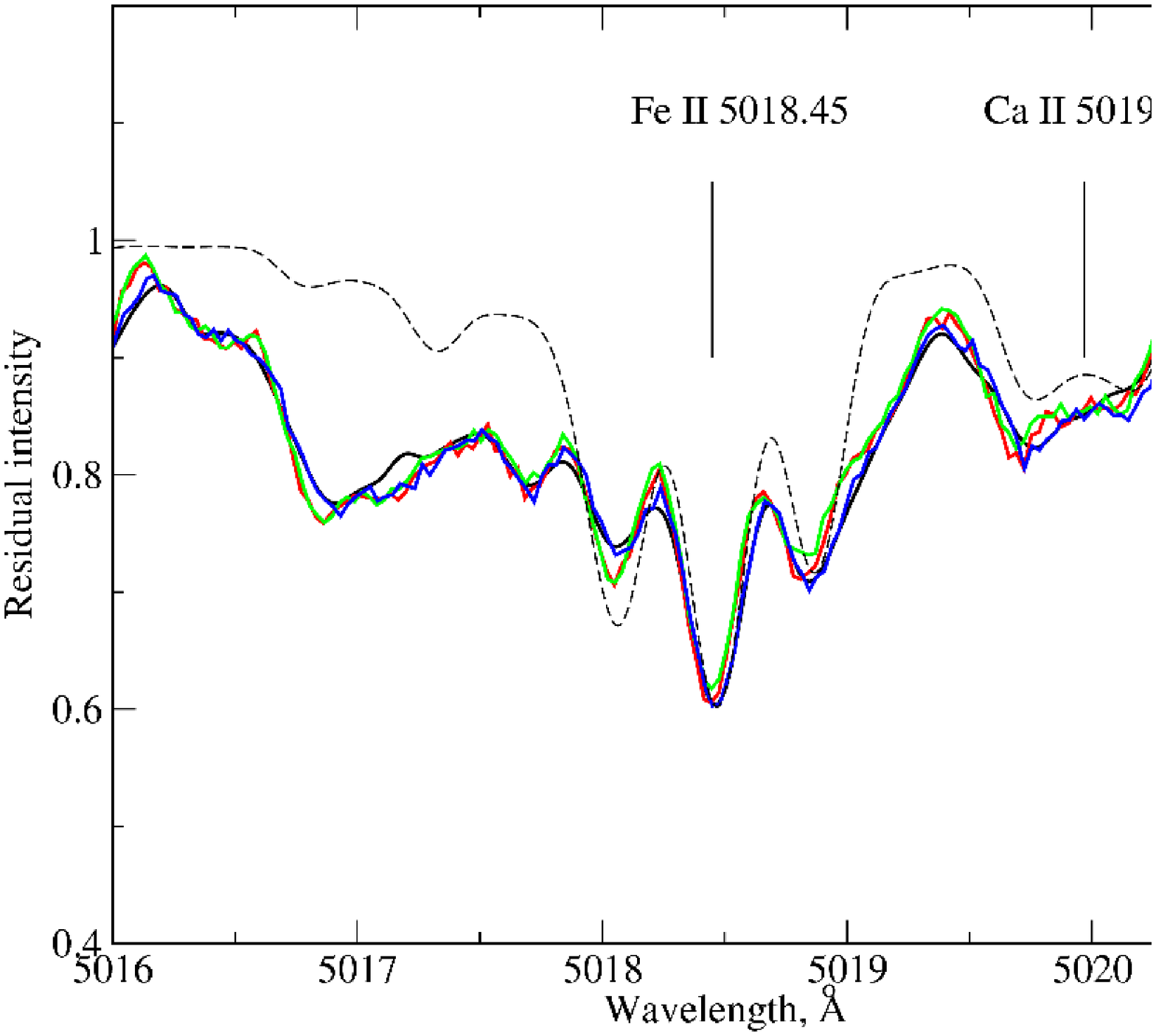}
\fifps{7.3cm}{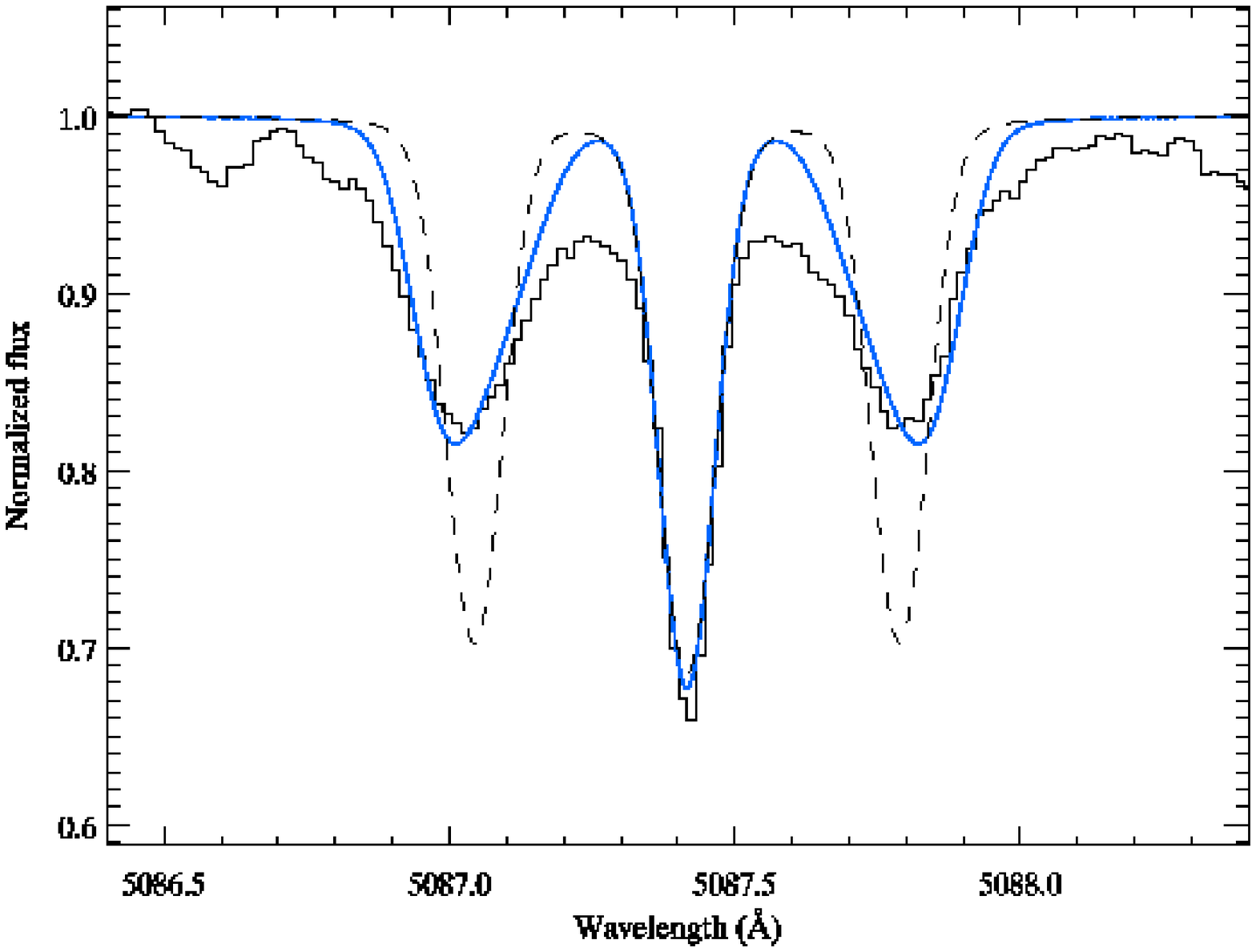}
\caption{Examples of the \sm\ modelling of the spectra of magnetic CP stars.
{\it Top panel}: comparison between observations of the red Si~{\sc ii} doublet at $\lambda$
6347 and 6371~\AA\ and 
spectrum synthesis for HD\,137509 --
the star with second-largest field known (\bsv{29}, Kochukhov 2006). {\it Bottom
left panel}: detection of the Zeeman resolved components in the Fe~{\sc ii}
5018~\AA\ line in the spectrum of cool Ap star HD\,178892 (Ryabchikova et al.
2006a). In this plot solid lines show observations obtained with the NES
instrument at the SAO 6-m telescope, whereas dashed line shows \sm\ calculation
for \bsv{17.5}. {\it Bottom right panel}: observations (histogram) of the Y~{\sc ii} 
5087~\AA\ line in the strongly magnetic cool Ap star HD\,154708. Usual \sm\
calculation for \bsv{24} (dashed line) is compared with the result of combining
spectra for a dipolar-like distribution of the field strength in the range from 29 to 
16.5~kG (solid line).}
\label{fig3}
\end{figure*}

Early attempts to deduce vertical chemical stratification in the atmospheres of Ap stars were
based on the trial-and-error method (Ryabchikova et al. 2002). Abundances in the upper and lower
parts of the stellar atmosphere were manually adjusted to fit the line profile shapes and to remove
a discrepancy between the strength of lines of different excitation potential, equivalent width and
ionization stages. We have recently developed an automatic procedure, {\tt DDAFit}, to find
chemical abundance gradients from observed stellar spectra (Ryabchikova et al. 2005). The {\tt
DDAFit} script is written in IDL and provides an optimization and visualization interface to
the spectrum synthesis calculations with \sm\ and \st. Vertical abundance distributions are
described with the four parameters: chemical abundance in the upper atmosphere, abundance in
deep layers, the vertical position of abundance step and the width of the transition region
where chemical abundance changes between the two values. All four parameters can be optimized
simultaneously with the Levenberg-Marquardt least-squares fitting routine (Bevington \& Robinson 1992)
and based on observations of unlimited number of spectral regions, possibly using different
weights in accordance to the quality or relative importance of the observations of particular
spectral features. The program derives one chemical stratification profile at a time, but is
able to account for any number of fixed stratified abundances. 

\section{Applications}

\subsection{Magnetic stars}

\sm\ was used in a number of recent studies of magnetic Ap stars. In such analyses polarized
radiative transfer calculations are essential to derive accurate abundances, resolve contributors
to complex blends and to measure magnetic field strength.

Fig.~3 illustrates application of \sm\ to three different chemically peculiar stars with strong
magnetic field. For the Si-peculiar Bp star HD\,137509 (Kochukhov 2006) we show comparison
of observations and theoretical spectra in the region of the \ion{Si}{ii} doublet at
$\lambda$ 6347 and 6371~\AA. Strikingly different profile shapes of the two, otherwise 
similar, lines can be traced to the different Zeeman splitting structure. Resolved Zeeman components in
the \ion{Si}{ii} 6371~\AA\ line indicate \bsv{29}, which is a second-largest field ever
observed in a non-degenerate star. This very large field strength is confirmed by the spectrum synthesis
of other metal lines.

We also show application of \sm\ to the analysis of HD\,178892 -- a star discovered by the
magnetic surveys at the SAO 6-m telescope (Elkin et al. 2002). The follow up high-resolution
spectroscopy with NES (Ryabchikova et al. 2006a) reveal resolved Zeeman components in many
lines, for example \ion{Fe}{ii} 5018~\AA\ shown in Fig.~3. Low \te, strong magnetic field
(\bsv{17.5}) and relatively short rotation period make HD\,178892 an excellent candidate for
a detailed analysis of the horizontal structure of magnetic field and chemical abundance spots.

HD\,154708 is another unique cool Ap star, showing very strong magnetic field (\bsv{24.5},
Hubrig et al. 2005). Recent time-series observations at ESO suggest the presence of
low-amplitude $p$-mode pulsations in this star (Kurtz et al. 2006). Due to an extremely slow
rotation, Zeeman components are fully resolved for many spectral lines in HD\,154708. The field
is so strong however, that the standard \sm\ approximation of the surface field strength 
distribution with a single value of the field modulus fails (see Fig.~3). This is evident from the
comparison of spectrum synthesis and observed profiles of the $\sigma$ components in the 
Zeeman split triplet lines: $\sigma$ components are broader and shallower in observations, indicating a
non-negligible surface scatter of the field strength. This situation can be addressed with a
straightforward extension of the spectrum synthesis with \sm. Fig.~3 shows that a much better
agreement with observations can be achieved if the stellar flux profiles are produced from a
set of \sm\ computations for different field strengths, corresponding to the $B_{\rm p}=30$~kG
dipolar field topology.

\begin{figure*}[t]
\fifps{14cm}{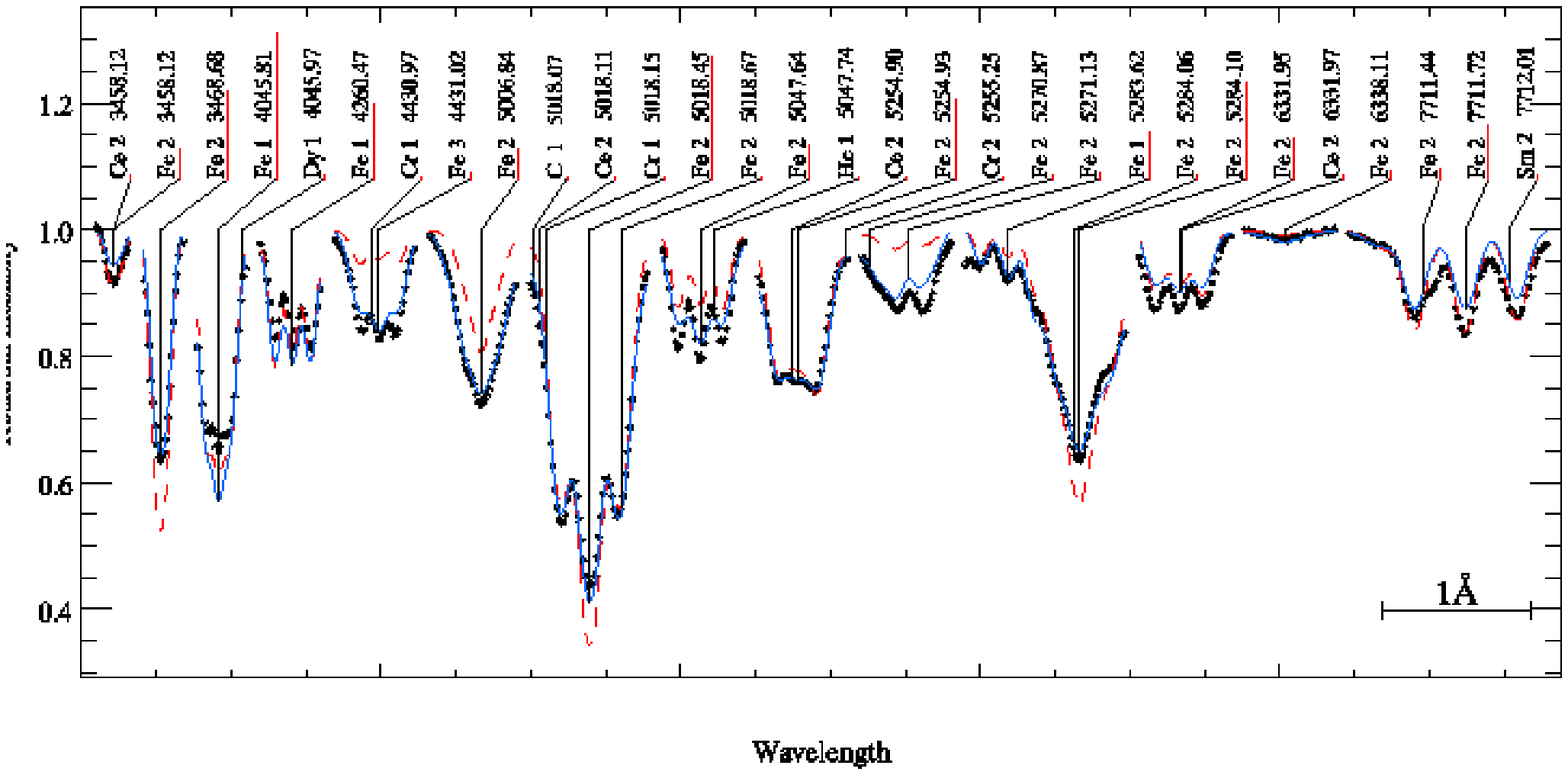}
\caption{Spectrum synthesis calculation for chemically stratified atmosphere
of the Ap star HD\,144897 (Ryabchikova et al. 2006b). Symbols show observed
line profiles. Dashed curve illustrates theoretical spectrum for constant Fe
abundance. A better fit to observations (solid line) could be achieved by taking into account
vertical stratification of Fe.}
\label{fig4}
\end{figure*}

\subsection{Chemically stratified atmospheres}

Both \sm\ and \st\ can calculate spectra for the prescribed inhomogeneous vertical distribution
of chemical elements. Parameters of the simple step-function distribution of abundances with depth
can be adjusted with the help of the {\tt DDAFit} script described above. 

A number of recent chemical abundance analyses of Ap stars accounted for chemical stratification in
the spectrum synthesis with \sm\ and \st\ and obtained vertical distributions of selected elements
using the {\tt DDAFit} procedure. Fig.~4 illustrates application of \sm\ to the problem of deriving
Fe stratification in the strongly magnetic Ap star HD\,144897 (Ryabchikova et al. 2006b). Another
example of employing \st\ and {\tt DDAFit} to the chemical stratification modelling can be found in Ryabchikova et al.
(2005). 

\section{Conclusions}

We have developed a new family of spectrum synthesis codes for modelling the spectra of stars with
various atmospheric parameters and chemistry. The codes \sm\ and \st\ can be used to calculate line
profiles of normal stars with different \te, as well as peculiar stellar objects with strong magnetic
fields, non-solar chemical composition and vertical chemical abundance gradients. Inclusion of the
molecular opacities and employment of the general equation of state solver permits application of our
codes to a wide variety of stars across the H-R diagramm. 

\begin{acknowledgements}
I thank N. Piskunov, T. Ryabchikova and P. Barklem for supplying various 
essential routines and atomic data included in the \st\ and \sm\ codes. 

My participation in the \textit{Magnetic Stars 2006} meeting at SAO was 
supported by the grants from the Swedish Kungliga Fysiografiska S\"allskapet 
and the Royal Academy of Sciences.
\end{acknowledgements}

\end{document}